\def\et{{\rm et al}~}
\documentstyle{mn}
\title[MONTE CARLO SIMULATIONS OF STAR CLUSTERS II]
{Monte Carlo simulations of star clusters -- II. Tidally limited,
multi--mass systems with stellar evolution.}
\author[Mirek Giersz]{Mirek Giersz$\rm ^{1}$ \\
$\rm ^1$ N. Copernicus Astronomical Center, Polish Academy of Sciences, ul.
Bartycka 18, 00--716 Warsaw, Poland\\}

\date{\small Accepted 2000- --- --. Received 2000- --- --; in original form
2000-  --- --}

\input epsf.sty

\begin{document}

\maketitle

\begin{abstract}

A revision of Stod\'o\l kiewicz's Monte Carlo code is used to
simulate evolution of large star clusters. The new method treats
each {\it superstar} as a single star and follows the evolution
and motion of all individual stellar objects. A survey of the
evolution of $N$--body systems influenced by the tidal field of a
parent galaxy and by stellar evolution is presented. The process
of energy generation is realised by means of appropriately
modified versions of Spitzer's and Mikkola's formulae for the interaction
cross section between binaries and field stars and binaries
themselves. The results presented  are in good agreement with
theoretical expectations and the results of other methods
(Fokker--Planck, Monte Carlo and $N$--body). The initial rapid
mass loss, due to stellar evolution of the most massive stars, causes
expansion of the whole cluster and eventually leads to the
disruption of less bound systems ($W_0=3$). Models with larger
$W_0$ survive this phase of evolution and then undergo core
collapse and subsequent post-collapse expansion, like isolated
models. The expansion phase is eventually reversed when tidal
limitation becomes important. The results presented are the first
major step in the direction of simulating evolution of real
globular clusters by means of the Monte Carlo method.

\end{abstract}

\vspace{1.0cm}

\begin{keywords}
\quad globular clusters: general \quad --- \quad methods: numerical \quad ---
\quad stars: kinematics
\end{keywords}

\section{INTRODUCTION.}

Understanding the dynamical evolution of self--graviting stellar
systems is one of the grand challenges of astrophysics. Among its
many applications, the one which motivates the work presented here
is the study of real globular clusters. Unfortunately, dynamical
modelling of globular clusters and other large collisional stellar
systems (like galactic nuclei, rich open clusters and galaxy
clusters) still suffers from severe drawbacks. These are, on the
one hand, due to the very high (and presently unfulfilled)
hardware requirements needed to model realistic, large stellar
systems by use of the direct $N$--body method, and partly due to
the poor understanding of the validity of assumptions used in
statistical modeling based on the Fokker--Planck and other
approximations. The Fokker--Planck method with finite differences,
which has recently been greatly improved, can now be used to
simulate more realistic stellar systems. It can tackle:
anisotropy, rotation, a tidal boundary, tidal shocking by galactic
disk and bulge, a mass spectrum, stellar evolution and dynamical
and primordial binaries (Takahashi 1995, 1996, 1997, Takahashi \&
Portegies Zwart 1998, 1999 hereafter TPZ, Drukier {\it et al.}
1999, Einsel \& Spurzem 1999, Takahashi \& Lee 1999 ).
Unfortunately, the Fokker--Planck approach suffers, among other
things, from the uncertainty of differential cross--sections of
many processes which are important during cluster evolution. It
can not supply detailed information about the formation and
movement of all objects present in clusters. Additionally, the
introduction of the mass function, mass loss and tidal boundary
into the code is very approximate. Direct $N$--body codes are the
most natural methods that can be used in simulations of real star
clusters (NBODY -- Aarseth 1985, 1999a, 1999b, Spurzem \& Aarseth
1996, Portegies Zwart {\it et al.} 1998 -- KIRA). They practically
do not suffer from any restrictions common in the Fokker--Planck
method. But unfortunately, even special--purpose hardware (Makino
{\et al.} 1997 and references therein) can be used effectively
only in $N$--body simulations which are limited to a rather
unrealistic number of stars (i.e. too small for globular
clusters). Another possibility is to use a code which is very fast
and provides a clear and unambiguous way of introducing all the
physical processes which are important during globular cluster
evolution. Monte Carlo codes, which use a statistical method of
solving the Fokker--Planck equation, provide all the necessary
flexibility. They were developed by Spitzer (1975, and references
therein) and H\'enon (1975, and references therein) in the early
seventies, and substantially improved by Marchant \& Shapiro
(1980, and references therein) and Stod\'o{\l}kiewicz (1986, and
references therein) and recently reintroduced by Giersz (1998,
hereafter Paper I, Giersz 2000, see also Giersz in Heggie {\it et
al.} 1999) and Joshi {\it et al.} (1999a), Joshi {\it et al.}
(1999b, hereafter JNR). The Monte Carlo scheme takes full
advantage of the established physical knowledge about the secular
evolution of (spherical) star clusters as inferred from continuum
model simulations. Additionally, it describes in a proper way the
graininess of the gravitational field and the stochasticity of
real $N$--body systems and provides, in a manner as detailed as in
direct $N$--body simulations, information about the movement of
any objects in the system. This does not include any additional
physical approximations or assumptions which are common in
Fokker--Planck and gas models (e.g. for conductivity). Because of
this, the Monte Carlo scheme can be regarded as a method which
lies between direct $N$--body and Fokker--Planck models and
combines most of their advantages.

Very detailed observations of globular clusters have extended our
knowledge about their stellar content, internal dynamics and the influence
of the environment on cluster evolution (Janes 1991, Djorgovski \&
Meylan 1993, Smith \& Brodie 1993, Hut \& Makino 1996, Meylan \&
Heggie 1997). They suggest that the galactic environment has
a very significant effect on cluster evolution. Gnedin {\it et al} (1999)
showed that, in addition to the well known importance of gravitational shock
heating of the cluster due to passages through the disk or close to the bulge
"shock--induced relaxation" (the second order perturbation) also has a
crucial influence on the cluster destruction rate. A
wealth of information on "peculiar" objects in globular clusters
(blue stragglers, X--ray sources (high-- and low--luminosity),
millisecond pulsars, etc.)  suggests a very close interplay between
stellar evolution, binary evolution and dynamical interactions.
This interplay is far from being understood. Moreover, recent
observations suggest that the primordial binary fraction in a
globular cluster can be as high as $15\%$ -- $38\%$ (Rubenstein \&
Bailyn 1997). Monte Carlo codes provide all the necessary
flexibility to disentangle the mutual interactions between all physical
processes which are important during globular cluster evolution.

The ultimate aim of the project described here is to build a Monte
Carlo code which will be able to simulate the evolution of real globular
clusters, as closely as possible. In this paper (the second in the
series) the Monte Carlo code (which was discussed in detail in Paper I) is
extended to include stellar evolution (as described by Chernoff \&
Weinberg (1990), hereafter CW or Tout {\it et al.} 1997),
multi--component systems (simulated by continuous, power--law mass
function), the tidal field of the Galaxy (simulated by a tidal radius and
an appropriate escape criterion) and binary--binary energy generation
(as introduced by Stod\'o{\l}kiewicz 1986). This is the first
major step in the direction of performing numerical simulations of
the dynamical evolution of real globular clusters submerged in the
Galaxy potential. The results of simulations will be compared with
those of CW, Aarseth \& Heggie (1998, hereafter AH) TPZ and JNR.

The plan of the paper is as follows. In Section 2, a short
description of the new features introduced into the Monte Carlo
code will be presented. In Section 3, the initial conditions will
be discussed and results of the simulation will be  shown. And
finally, in Section 4 the conclusions will be presented.

\section{MONTE CARLO METHOD.}

The Monte Carlo method can be regarded as a statistical way of
solving the Fokker--Planck equation. Its implementation presented
in Paper I is based on the orbit-averaged Monte Carlo method
developed in the early seventies by H\'enon (1971) and then
substantially improved by Stod\'o{\l}kiewicz (1986, and references
therein), Giersz (1998) and recently by JNR. The code is described
in detail in Paper I, which deals with simulations of isolated,
single--mass systems. In this section, additional physical
processes (not included in Paper I) will be discussed: multi--mass
systems and stellar evolution (\S 2.1), mass loss through the
tidal boundary (\S 2.2), formation of three--body binaries and
their subsequent interactions with field stars in multi--component
systems (\S 2.3), and interactions between binaries (\S 2.4). The
implementation of an arbitrary mass spectrum in the Monte Carlo
method is very straightforward and will not be discussed here
(see for example Stod\'o{\l}kiewicz 1982).

\subsection{Mass Spectrum and Stellar Evolution}

Observations give more and more evidence that the mass function
in globular/open clusters and for field stars as well is not a
simple power--law, but is rather approximated by a composite
power--law (e.g. Kroupa, Tout \& Gilmore 1993). However, in the
present study the simple power--law is used, for simplicity and to
facilitate comparison with other numerical simulations (CW,
Fukushige \& Heggie 1995, Giersz \& Heggie 1997, AH, TPZ, JNR).
The following form for the initial mass function was assumed:

\begin{equation}
N(m)dm = Cm^{-\alpha}dm,   \quad\quad\quad  m_{min} \leq m \leq
m_{max},
\end{equation}
where C and $\alpha$ are constants. The detailed discussion of
$\alpha$, $m_{min}$ and $m_{max}$ will be presented in \S 3.1. To
describe the mass loss due to stellar evolution
the same simplified stellar evolution model as adopted by CW was
used. More sophisticated models (Portegies Zwart \& Verbunt 1996,
Tout {\it et al.} 1997) give very similar results for the time which
a star spends on the main--sequence, and also slightly smaller masses
of remnants of stellar evolution (see Table 1). \begin{table}
\begin{center}
\caption{Main--sequence lifetimes and remnant masses $^a$}
\begin{tabular}{|c|c|c|c|c|} \hline\hline
  $m_{initial}$ & $log(t_{MS})$  & $log(t_{MS})$  & $m_{remnant}$  &
$m_{remnant}$  \\ \\
   \cline{2-5} \\
  \quad & CW  & T  & CW  & T  \\ \hline
  0.40  & 11.30  & 11.26  & 0.45  & 0.39  \\
  0.60  & 10.70  & 10.73  & 0.49  & 0.41  \\
  0.80  & 10.20  & 10.28  & 0.54  & 0.52  \\
  1.00  & 9.89  & 9.92  & 0.58  & 0.57  \\
  2.00  & 8,80  & 8.89  & 0.80  & 0.75  \\
  4.00  & 7.95  & 8.13  & 1.24  & 1.09  \\
  8.00  & 7.34  & 7.52  & 0.00  & 0.00  \\
  15.00  & 6.93  & 7.06  & 1.40  & 1.40  \\ \hline
  \multicolumn{5}{l}{} \\
  \multicolumn{5}{l}{$^a$ Masses are in units of the Solar mass. Main--sequence lifetime is}\\
  \multicolumn{5}{l}{in years. Columns labeled by CW and T give data from}\\
  \multicolumn{5}{l}{Chernoff \& Weinberg (1990) and Tout {\it et al.} (1997),
  respectively.}\\
\end{tabular}
\end{center}
\end{table}
This should not significantly change the results of simulations.
In CW it is assumed that a star at the end of its main--sequence
lifetime instantaneously ejects its envelope and becomes a compact
remnant (white dwarf, neutron star or black hole). This is a good
approximation, since (from the point of view of cluster evolution)
the dominant mass loss phase occurs on a time scale shorter than a
few Myrs, and this is much less than the cluster evolution time
scale (proportional to the relaxation time), which is of the order
1 Gyr. Mass loss due to stellar winds is neglected for
main--sequence  stars. According to the prescription given in CW;
main--sequence stars of mass $m > 8 M_{\odot}$ finish their
evolution as neutron stars of mass $1.4 M_{\odot}$, while stars of
mass $m < 4 M_{\odot}$ end as white dwarfs of mass $0.58 + 0.22(m
- 1)$. Stars of intermediate masses are completely destroyed. For
stars with masses smaller than $\simeq 0.83 M_{\odot}$ the
main--sequence lifetime is linearly (log -- log) extrapolated,
which is in very good agreement with the scaling $m^{-3.5}$ used
by TPZ and JNR. The initial masses of stars are generated from the
continuous distribution given in equation (1). This ensures a
natural spread in their lifetimes. To ensures that the cluster
remains close to virial equilibrium during rapid mass loss due to
stellar evolution, special care is taken that no more than $3\%$
of the total cluster mass is lost during one overall time--step.

\subsection{Tidal Stripping}

For a tidally truncated cluster the mass loss from the system is
dominated by tidal stripping -- diffusion across the tidal
boundary. This leads to a much higher rate of mass loss than in an
isolated system, where mass loss is attributed mainly to
rare strong interactions in the dense, inner part of the system.
In the present Monte Carlo code, a mixed criterion is used to
identify escapers: a combination of apocenter and energy--based criteria.
In the apocenter criterion a star is removed from the system if

\begin{equation}
r_a(E,J) > r_t,
\end{equation}
and in the energy criterion a star is removed if

\begin{equation}
E > E_t \equiv -GM/r_t,
\end{equation}
where $r_a(E,J)$ is the apocenter distance of a star with energy
$E$ and angular momentum $J$, $r_t$ is the tidal radius of the
cluster of mass $M$, and $G$ is the gravitational constant. Recently,
TPZ demonstrated that the energy--based criterion can lead to an
overestimation of the escape rate from a cluster. Sometimes
stars on nearly circular orbits well inside the tidal radius can be
removed from the system . The use of the mixed criterion can even
further increase the escape rate and shorten the time to cluster
disruption. No potential escapers are kept in the system. This mixed
criterion was mainly used to facilitate direct comparison with
Monte Carlo simulations presented in the collaborative experiment
(Heggie {\it et al.} 1999). Stars regarded as escapers are lost
instantaneously from the system. This is in contrast to $N$--body
simulations where stars need time proportional to the dynamical
time to be removed from the system. Recently, Baumgardt (2000)
showed that stars with energy greater than $E_t$ in the course of
escape can again become bound to the system, because of distant
interactions with field stars. This process can substantially
influence the escape rate. The mass loss across the tidal boundary
can become unstable, when too many stars are removed from the system
at the same time.  This is characteristic of the final
stages of cluster evolution, and for clusters with initially
low central concentration. To properly follow these stages of
evolution the time--step has to be decreased to force smaller mass
loss. JNR used an iteration procedure to determine the mass loss
and the tidal radius. A similar procedure will be introduced
into a future version of our Monte Carlo code.

\subsection{Three--Body Binaries}

In the present Monte Carlo code (as it was described in Paper I)
all stellar objects, including binaries, are treated as single
{\it superstars}. This allows one to introduce into the code, in a
simple and accurate way, processes of stochastic formation of
binaries and their subsequent stochastic interactions with field
stars and other binaries. The procedure for single--component
stellar systems was described in detail in Paper I. Now the
procedure for multi--component systems will be discussed.

Suppose that the rate of formation, per unit volume, of
three-body binaries with components of mass $m_1$ and $m_2$, by
interaction with stars of mass $m_3$ is

\begin{equation}
{{dn_b}\over dt} = Bn_1n_2n_3,
\end{equation}
where $n_1$, $n_2$ and $n_3$ are the number densities of stars of mass
$m_1$, $m_2$ and $m_3$, respectively. The coefficient B is a
function of the masses of the interacting stars and the local mean kinetic
energy (see for example Heggie 1975 and Stod\'o{\l}kiewicz 1986).
Suppose also, that the formation of binaries within a radial zone
(superzone, see Paper I) of volume $\Delta V$ in a time interval
$\Delta t$ is considered. The number of binaries with components
of mass $m_1$, $m_2$, formed by interaction with stars of mass
$m_3$, in this zone, in this time interval is

\begin{equation}
\Delta n_b = Bn_1n_2n_3\Delta V \Delta t.
\end{equation}
Now suppose that all stars in this zone are divided into groups
of three, and for each group a binary is created from the first
two stars in the group with probability $P$. Suppose there are
$\Delta N$ shells in the zone. Then, there are $\Delta N/3$ groups
of three stars. Also, the probability that, in one given group,
the masses of the three stars are $m_1$, $m_2$ and $m_3$ (as
required), is $n_1n_2n_3/n^3$ (where $n$ is the total number density).
Therefore, the average number of binaries formed in this zone in
this time interval, with components of mass $m_1$ and $m_2$ (by
interaction with a star of mass $m_3$) is

\begin{equation}
\Delta n_b = {\Delta N\over 3}{n_1n_2n_3\over n^3}P.
\end{equation}
This is equal to the required number (equation 5) if

\begin{equation}
P = {{3Bn^3\Delta V\Delta t}\over \Delta N}.
\end{equation}
So, to determine the probability of the formation of a three--body
binary with components of mass $m_1$, $m_2$ and $m_3$ one has to
know only the local total number density instead of the local number
density for stars of each mass. This procedure substantially
reduces fluctuations in the binary formation rate. The determination of the
local number density in the Monte Carlo code is a very delicate matter
(Paper I). It is practically impossible if one has to
compute the local number densities for each species of a
multi--component system.

A binary living in the cluster is influenced by close and wide
interactions with field stars. Close interactions are the most
important from the point of view of cluster evolution. They
generate energy, which supports post--collapse cluster evolution.
For a single--component system the probability of a binary field
star interaction can be computed using the well known simple
semi--empirical formula of Spitzer (Spitzer 1987). For
multi--component systems there is no simple semi--empirical
formula which can fit all numerical data (Heggie {\it et al.}
1996). In the present Monte Carlo model, the following strategy
was used to compute the total probability of interaction between a
binary of binding energy $E_b$ consisting of mass $m_1$ and $m_2$
and a field star of mass $m_3$. The mass dependence of the
probability was deduced from equations (6-23), (6-11) and (6-12)
presented in Spitzer (1987) and results of Heggie (1975). Then, the
coefficient was suitably adjusted, so that the total probability
was correct for the equal--mass case. The formula obtained in such a
way is as follows

\begin{equation}
P_{3b*} = \int{5\sqrt2\pi A_sG^2m_1^2m_2^2\sqrt
{m_{123}}n\over{8\sqrt3\sqrt{m_{12}}\sqrt{m_3} \sqrt{m_a} \sigma
E_b}}dt,
\end{equation}
where $m_{12} = m_1 + m_2$, $m_{123} = m_{12} + m_3$, $m_a$ is an
average stellar mass in a zone, and $\sigma$ is the one dimensional
velocity dispersion. This procedure is, of course, oversimplified
and in some situations can not give correct results, for example,
when a field star is very light compared to the mass of the binary
components. The changes of the binding energy of binaries and their
velocities due to interactions with field stars are computed in
the same way as in Paper I.

\subsection{Binary--Binary interactions}

It is well known that interactions between binaries can play an
important role in globular cluster evolution, particularly when
primordial binaries are present (Gao {\it et al.} 1991, Hut {\it
et al.} 1992 and reference therein, Meylan \& Heggie 1998 and
reference therein, Giersz \& Spurzem 2000). Binary--binary
interactions, besides their role in energy generation, can also be
involved in the creation of many different peculiar
objects observed in globular clusters. The problem of energy
generation in binary--binary interactions is very difficult to
solve (see the pioneering work of Mikkola (1983, 1984)).
The implementation of interactions between binaries in the Monte
Carlo code is based on the method described by Stod\'o{\l}kiewicz
(1986). Only strong interactions are considered, and only two
types of outcomes are permitted: one binary (composed of the
heaviest components of the interacting binaries) and two single stars,
or two binaries in a hyperbolic relative orbit. The eccentricities of
the orbits of both binaries, and their orientations in space, are
neglected. Also stable three--body configurations are not allowed.

In the case when binaries are formed in dynamical processes, and
only a few binaries are present at any time in the system, it is
very difficult to use the binary density (Giersz \& Spurzem 2000) to
calculate the probability of a binary--binary interaction. Another
approach has to be employed (Stod\'o{\l}kiewicz 1986). For a given
binary, the pericenter, $r_-$, and apocenter, $r_+$, distances of
its orbit in the cluster are known. This binary can hit only
binaries (regarded as targets) whose actual distance from the
cluster center lie between $r_-$ and $r_+$. To compute the
probability of this interaction, the following procedure is
introduced (a modification of the procedure proposed by
Stod\'o{\l}kiewicz 1986). The rate of encounters of binaries
within impact parameter $p$ is

\begin{equation}
\int f({\b v_1})n_1({\b r})f({\b v_2})n_2({\b r})|{\b v_1} - {\b
v_2}|\pi p^2 d^3{\b v_1}d^3{\b v_2}d^3{\b r},
\end{equation}
where $f$ is a distribution function normalized to unity, $n_1$
and $n_2$ are the number densities of binaries, and ${\b v_1}$ and ${\b
v_2}$ are their barycentric velocities. Suppose there is a binary
at a given position $\b r$ with velocity ${\b v^\prime}$. The rate
of encounters with other binaries at an impact parameter less than
$p$ is

\begin{equation}
\dot{N} = \int f({\b v})n({\b r})|{\b v} - {\b v^\prime}|\pi p^2
d^3{\b v},
\end{equation}
where $f$ is normalized to unity and $n$ is normalized to the
total number of binaries. The number of encounters in time $\Delta
t$ is

\begin{equation}
\Delta N = \Delta t \int f({\b v})n({\b r})|{\b v} - {\b
v^\prime}|\pi p^2 d^3{\b v}.
\end{equation}
The number of encounters with one specified binary is

\begin{equation}
\Delta N_1 = \Delta t \int f({\b v})n_s({\b r})|{\b v} - {\b
v^\prime}|\pi p^2 d^3{\b v},
\end{equation}
where $n_s$ is normalized to unity. Equation 12 can be written in
the form

\begin{equation}
<\Delta N_1> = n_s({\b r})<|{\b v} - {\b v^\prime}|>\pi p^2\Delta
t.
\end{equation}
Now, $n_s(\b r)4\pi r^2$ is the probability density of $r$, i.e.

\begin{equation}
4\pi r^2 n_s({\b r})dr = \left\{
\begin{array}{ccl}
{{dr/|\dot r|}\over{\int\limits_{r_-}^{r_+}{dr\over{\dot r}}}} &
for & r_-< r <r_+ \\ 0 & \quad & otherwise
\end{array}
\right.
\end{equation}
So, $4\pi r^2 n_s({\b r}) = 2/|\dot r|/T$, where $T$ is the orbital
period of the binary in the cluster. Hence

\begin{equation}
n_s(\b r) = {1 \over {2\pi r^2 T |\dot{r}|}}.
\end{equation}
Substituting equation (15) into equation (13) we may obtain a formula for the
probability $P_{3b3b}$ of the encounter between the two binaries, i.e.

\begin{equation}
P_{3b3b} = {w\over |v_r|} {p^2\over {2r^2T}} \Delta t,
\end{equation}
where $w = <|\b v - \b v^\prime|>$, and $|v_r| = |\dot{r}|$. It is
interesting to note that equation (16) is very similar to
equation (36) in Stod\'o{\l}kiewicz (1986), which was obtained in
a very approximate way. It differs only by the factor $w/|v_r|$, which
takes into account in a proper way the geometry of the encounter. To
compute the maximum impact parameter, a value equal to $2.5$ times
the semi--major axis of the softer binary (according to
Stod\'o{\l}kiewicz's (1986) prescription) for the minimum distance
during the encounter was adopted. The probability $P_{3b3b}$ is
evaluated every time step for each pair of binaries (when binaries are
sorted against increasing distance from the cluster center). If
this probability is smaller than a random number (drawn from a
uniform distribution) the binaries are due to interact. The outcome of
the interaction is as follows (see details in Stod\'o{\l}kiewicz
1986):
\begin{itemize}
  \item Two binaries in a hyperbolic relative orbit - $12\%$ of
  all interactions. In this case it is assumed that the recoil
  energy received by both binaries is equal to $0.4$ times the binding
  energy of the softer binary, and this binding energy increases
  by the same value.
  \item One binary and two escapers - $88\%$ of all interactions.
  The total recoil energy released is equal to $0.516(E_{b1} +
  E_{b2})$ and distributed according to conservation of momentum.
  The softer binary is destroyed and the harder binary increases its
  binding energy by the amount equal to the recoil energy. $E_{b1}$
  and $E_{b2}$ are the binding energies of interacting binaries.
\end{itemize}
The new values of the velocity components of the binaries/singles are
computed as discussed in Stod\'o{\l}kiewicz (1986) and Giersz
(1998).

The procedure described above is very uncertain in regard to the
amount of energy generated by binaries in their interactions with
field stars and other binaries  (particularly for multi--component
systems). To solve this problem, it is planned in future to introduce into the
Monte Carlo code numerical procedures (based on Aarseth's
NBODY6 code) which can numerically integrate the motion of three-- and
four--body subsystems. This will at least ensure that the
energy generated in these interactions will be calculated properly.
A similar procedure was introduced, with success, into the Hybrid
code (Giersz \& Spurzem 2000). But still there will be remaining
uncertainties in the determination of the overall probabilities
for binary creation and binary--binary interactions.

\section{RESULTS}

In Paper I the first results of Monte Carlo simulations of the
evolution of single--component systems were presented. Here, the
Monte Carlo code is extended to include a power--law mass
function, stellar evolution, tidal stripping and binary--binary
interactions.

\subsection{Initial Models}

The initial conditions were chosen in a similar way as in the
``collaborative experiment'' (Heggie {\it et al.} 1999). The positions
and velocities of all stars were drawn from a King model with a
power--law mass spectrum. All standard models have the same total
mass $M = 60000 M_{\odot}$ and the same tidal radius $R_c = 30$
pc. Masses are drawn from the power--law mass function according
to equation (1). The minimum mass was chosen to be $0.1 M_\odot$ and
the maximum mass be $15 M_\odot$. Three different values of the
power--law index were chosen: $\alpha = 1.5$, $2.35$ and $3.5$.
The sets of initial King models were characterized by $W_0 = 3$,
$5$ and $7$. To facilitate comparison with results of CW, AH, TPZ
and JNR additional models of CW's Family 1 were computed (minimum
mass equal to $0.4 M_\odot$, $\alpha = 1.5$, $2.5$, $3.5$ and
total mass $M = 90685 M_{\odot}$, $99100 M_{\odot}$, $103040
M_{\odot}$, respectively). In this paper the results of the
standard models will mainly be presented. Models of Family 1 will not
be shown (with one exception - Figure 12) and will only be discussed in
cases, where their results are different from these of the standard
models. All models are listed in Table 2.

\begin{table}
\begin{center}
\caption{Models $^a$}
\begin{tabular}{|c|c|c|c|c|c|} \hline \hline
 Model \quad & $W_0$ \quad & \quad $\alpha$ \quad & \quad $N_T$ \quad  & \quad $r_{t_M}$ \quad &$t_{scale}$\\ \hline
  W3235$^b$ & 3 & 2.35 & 187908 & 3.1311 & 34139 \\
  W335$^b$ & 3 & 3.5 & 360195 & 3.1311 & 61418 \\
  W515$^b$ & 5 & 1.5 & 48990 & 4.3576 & 6269 \\
  W5235$^b$ & 5 & 2.35 & 187908 & 4.3576 & 20793 \\
  W535$^b$ & 5 & 3.5 & 360195 & 4.3576 & 37409 \\
  W715$^b$ & 7 & 1.5 & 48990 & 6.9752 & 3096 \\
  W7235$^b$ & 7 & 2.35 & 187908 & 6.9752 & 10267 \\
  W735$^b$ & 7 & 3.5 & 360195 & 6.9752 & 18472 \\
  P10 & - & 2.35 & 187908 & 10.0 & 5982 \\
  W325-4$^c$ & 3 & 2.5 & 98217 & 3.1311 & 28490 \\
  W335-4$^c$ & 3 & 3.5 & 155218 & 3.1311 & 42902 \\
  W515-4$^c$ & 5 & 1.5 & 37022 & 4.3576 & 7309 \\
  W525-4$^c$ & 5 & 2.5 & 98217 & 4.3576 & 17361 \\
  W535-4$^c$ & 5 & 3.5 & 155218 & 4.3576 & 26131 \\
  W715-4$^c$ & 7 & 1.5 & 37022 & 6.9752 & 3609 \\
  W725-4$^c$ & 7 & 2.5 & 98271 & 6.9752 & 8573 \\
  W735-4$^c$ & 7 & 3.5 & 155218 & 6.9752 & 13872 \\ \hline
  \multicolumn{6}{l}{} \\
  \multicolumn{6}{l}{$^a$ $N_T$ is the total number of stars, $r_{t_M}$ is the tidal radius in}\\
  \multicolumn{6}{l}{Monte Carlo units and $t_{scale}$ is the time scaling factor to scale}\\
  \multicolumn{6}{l}{simulation time to physical time (in $10^6$ yrs).}\\
  \multicolumn{6}{l}{The first entry after W describes the King model and the}\\
  \multicolumn{6}{l}{following numbers the mass function power-law index.}\\
  \multicolumn{6}{l}{$P10$ is the Plummer model.}\\
  \multicolumn{6}{l}{$^b$ Standard models.}\\
  \multicolumn{6}{l}{$^c$ models of Family 1 (Chernoff \& Weinberg 1990).} \\
  \end{tabular}
\end{center}
\end{table}

The initial model is not in virial equilibrium, because of statistical noise
and because masses
are assigned independently from the positions and velocities. Therefore
it has to be initially rescaled to virial equilibrium. During all the
simulations the virial ratio is kept within $< 2\%$ of the equilibrium
value (for the worst
case within $<5\%$). Standard $N$--body units (Heggie \&
Mathieu 1986), in which the total mass $M = 1$, $G = 1$ and the initial total
energy of the cluster is equal to $-1/4$, have been adopted for all
runs. Monte Carlo time is equal to $N$--body time divided by
$N/ln(\gamma N)$, where gamma was adopted to be $0.11$ as for
single--component systems (Giersz \& Heggie 1994). However, there
are some results suggesting even smaller values of $\gamma$
($\approx 0.015$) for multi--components systems (Giersz \& Heggie
1996, 1997). In order to scale time to the physical units the
following formula is used

\begin{equation}
{t_{scale}\over 10^6 yrs} = 14.91\left({R_c\over
r_{t_M}}\right)^{1.5} {N_T \over {\sqrt{M} ln (\gamma N_T)}},
\end{equation}
where $r_{t_M}$ is the tidal radius in Monte Carlo units, $M$ and
$R_c$ are in {\it cgs} units. $r_{t_M}$ and $t_{scale}$ are listed in
Table 2. In the course of evolution, when the cluster loses mass, the
tidal radius changes according to $r_t \propto M^{1/3}$.

Finally, a few words about the efficiency of the Monte Carlo code are
presented here. The simulations of the largest models (consist of
$360195$ stars) took about one month on a Pentium II 300 MHz PC.
This is still much shorter than the biggest direct $N$-body
simulations performed on GRAPE--4 (Teraflop special--purpose
hardware). Nevertheless, the speed of the code is not high enough
to perform, in a reasonable time, a survey of models, as was done,
for example, by CW. It is clear, that to simulate evolution of real
globular clusters a substantial speed up of the code is needed.
This can be done, either by parallelizing the code (in a similar
way to JNR), introducing a more efficient way of determining the
new positions of {\it superstars}, or by using a hybrid code (as
was done by Giersz \& Spurzem 2000). However, the Monte
Carlo code has presently a great relative advantage over the $N$-body
code for simulations with a large number of primordial binaries.
Primordial binaries substantially downgrade the performance of
$N$-body codes on supercomputers or on special-purpose hardware.
They can be introduced into the Monte Carlo code in a natural way,
practically without a substantial loss of performance (Giersz \&
Spurzem 2000).

\subsection{Global evolution.}

To facilitate comparison of standard models with recently obtained
results of $N$--body (AH), Fokker--Planck (CW, TPZ) and Monte
Carlo (JNR) simulations of globular cluster evolution, the following
parameter (introduced by CW), which describes the initial
relaxation time, is calculated. It is defined by

\begin{equation}
F \equiv {M \over {M_{\odot}}} {R_g \over {kpc}} {220 km s^{-1}
\over {v_g}} {1 \over{lnN}},
\end{equation}
where $R_g$ is the distance of the globular cluster to the Galactic
center, and $v_g$ is the circular speed of the cluster around the
Galaxy. For the initial parameters of the standard models ($M =
60000 M_{\odot}$, $r_t = 30$ $pc$, $v_g = 220$ $km s^{-1}$), $R_g
= 3.89$ $kpc$. Therefore according to equation 18, $F$ is equal to
$2.14$x$10^4$, $1.92$x$10^4$ and $1.82$x$10^4$ for $\alpha$ equal
to $-1.5$, $-2.35$ and $-3.5$, respectively. This is about $2.5$
times smaller than for Family 1 of CW. Generally, the greater the value of $F$,
the longer the relaxation time and the slower the evolution. Therefore one
should expect, that for standard models the collapse or disruption
times are shorter than for CW, AH, TPZ or JNR. Additionally, because
the minimum stellar mass is $0.1 M_{\odot}$ instead of $0.4 M_{\odot}$
(the value adopted by CW, AH, TPZ and JNR), one should expect that
mass segregation will proceed faster in standard models, and the
collapse time should be further shortened. Indeed, as can be
seen in Table 3, this is true with the exception of models with a flat
mass function (W3235 and probably W715). For these models the
dominant physical process during most of the cluster life is
stellar evolution. The standard models contain a smaller number of
massive main--sequence stars than Family 1 models. Therefore they should
evolve more slowly.

The standard models show very good agreement with $N-$body results
(Heggie 2000). See columns labeled by G and H-0.1 in Table 3. Only
models with a flat mass function show substantial disagreement.
These models are difficult for both methods. Violent stellar
evolution and induced strong tidal stripping lead to troubles with
time--scaling for the $N-$body model and with the proper
determination of the tidal radius for the Monte Carlo model.
Generally, the same is true for Monte Carlo models of Family 1.
Results of these models show good agreement with the results of
CW, AH and TPZ. JNR's results, particularly for strongly
concentrated systems, disagree with all other models. This may be
connected with the fact that JNR's Monte Carlo scheme is not
particularly suitable for high central density and strong density
contrast. The deflection angles adopted by JNR are too large and
consequently the time-steps are too large, which can lead to too
fast evolution for these models.

\begin{table}
\begin{center}
\caption{Time of cluster collapse or disruption $^a$}
\begin{tabular}{|c|c|c|c|c|c|c|c|} \hline\hline
Model &CW&TPZ&JNR&AH&G-0.4&H-0.1&G\\ \hline
  W3235/25-4$^b$&0.28&2.2&5.2&2.1&0.7&11.3&6.3\\
  W335/35-4&21.5&32.0&31.0&$>$20.0&26.0&16.0&17.6\\
  W515/15-4$^b$&-&-&-&0.2&0.07&0.5&0.1\\
  W5235/25-4&-&-&-&13.5&13.2&7.0&6.8\\
  W535/35-4&-&-&-&$>$20.0&26.1&6.0&7.0\\
  W715/15-4$^b$&1.0&3.1&3.1&1.2&2.8&3.4&2.1\\
  W7235/25-4&9.6&10.0&3.0&11.0&9.8&1.7&1.9\\
  W735/35-4&10.5&9.9&6.0&9.2&10.7&0.8&0.7\\ \hline
  \multicolumn{8}{l}{} \\
  \multicolumn{8}{l}{$^a$ Time is given in $10^9$ yr.,} \\
  \multicolumn{8}{l}{In the column Model the entry before the slash
  is for}\\
  \multicolumn{8}{l}{standard models and after the slash for models of Family
  1}\\
  \multicolumn{8}{l}{CW --- Chernoff \& Weinberg (1990) --- Family 1,}\\
  \multicolumn{8}{l}{TPZ --- Takahashi \& Portegies Zwart (1999) --- Family 1,} \\
  \multicolumn{8}{l}{ JNR --- Joshi et al. (1999) --- Family 1,}\\
  \multicolumn{8}{l}{AH --- Aarseth \& Heggie (1998) --- Family 1,} \\
  \multicolumn{8}{l}{H-0.1 --- Heggie --- standard case --- $m_{min} = 0.1 M_{\odot}$,}\\
  \multicolumn{8}{l}{G --- Giersz --- standard case --- $m_{min} = 0.1 M_{\odot}$,}\\
  \multicolumn{8}{l}{ G-0.4 --- Giersz  --- Family 1,}\\
  \multicolumn{8}{l}{$^b$ Cluster was disrupted, other models collapsed.}\\
\end{tabular}
\end{center}
\end{table}

Figures 1 to 3 present the evolution of the total mass for
standard models.
\begin{figure}
\epsfverbosetrue
\begin{center}
\leavevmode \epsfxsize=80mm \epsfysize=60mm
\epsfbox{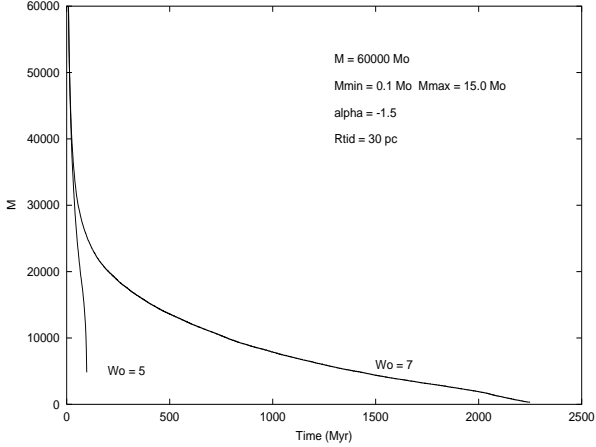}
\end{center}
\caption{Evolution of the total mass for models W515 and W715.}
\end{figure}
\begin{figure}
\epsfverbosetrue
\begin{center}
\leavevmode \epsfxsize=80mm \epsfysize=60mm
\epsfbox{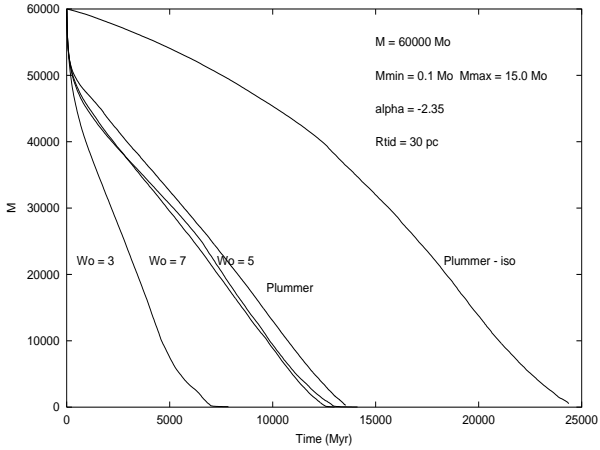}
\end{center}
\caption{Evolution of the total mass for models W3235, W5235,
W7235, P10 and an isolated Plummer model (Plummer--iso)}
\end{figure}
\begin{figure}
\epsfverbosetrue
\begin{center}
\leavevmode \epsfxsize=80mm \epsfysize=60mm
\epsfbox{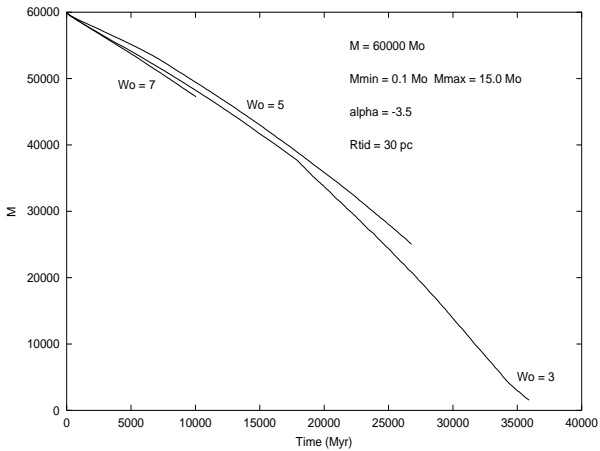}
\end{center}
\caption{Evolution of the total mass for models W335, W535 and
W735.}
\end{figure}
Three phases of evolution are clearly visible. The first
phase is connected with the violent mass loss due to
evolution of the most massive stars. Then there is the long phase
connected with gradual mass loss due to tidal stripping. And
finally there is the phase connected with the tidal disruption of a
cluster. This phase is not well represented by most models
presented here. The chosen overall time--step is too long to
follow in detail the evolution, which proceeds basically on a dynamical
time scale. Only for models W515 and W715 was a sufficiently small
time--step adopted to properly follow this phase of evolution.
Models with a steep mass--function ($\alpha \leq -2.35$) and with
different $W_o$ show very similar evolution during the phase of
tidal stripping (except W3235). It seems that the initial mass
loss due to stellar evolution is not sufficiently strong to
substantially change the initial cluster structure. Models of
Family 1 do not show this feature. It seems that the initial mass loss
across the tidal boundary is sufficiently strong to change the
structure of the system. These models contain initially a much larger
number of massive main--sequence stars than standard models. Figures 1
to 3 present qualitatively very similar features as figures shown by
TPZ and JNR.

Figures 4 to 6 show the amount of mass loss due to stellar
evolution, $evo$, and tidal stripping, $esc$, for standard models.
\begin{figure}
\epsfverbosetrue
\begin{center}
\leavevmode \epsfxsize=80mm \epsfysize=60mm
\epsfbox{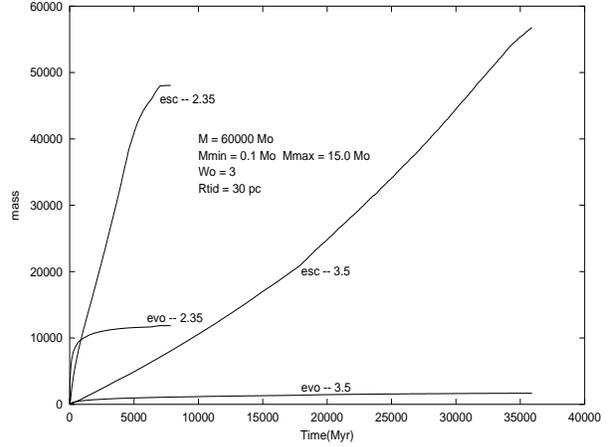}
\end{center}
\caption{Mass loss due to stellar evolution (evo) and tidal
stripping (esc) for models W3235 and W335.}
\end{figure}
\begin{figure}
\epsfverbosetrue
\begin{center}
\leavevmode \epsfxsize=80mm \epsfysize=60mm
\epsfbox{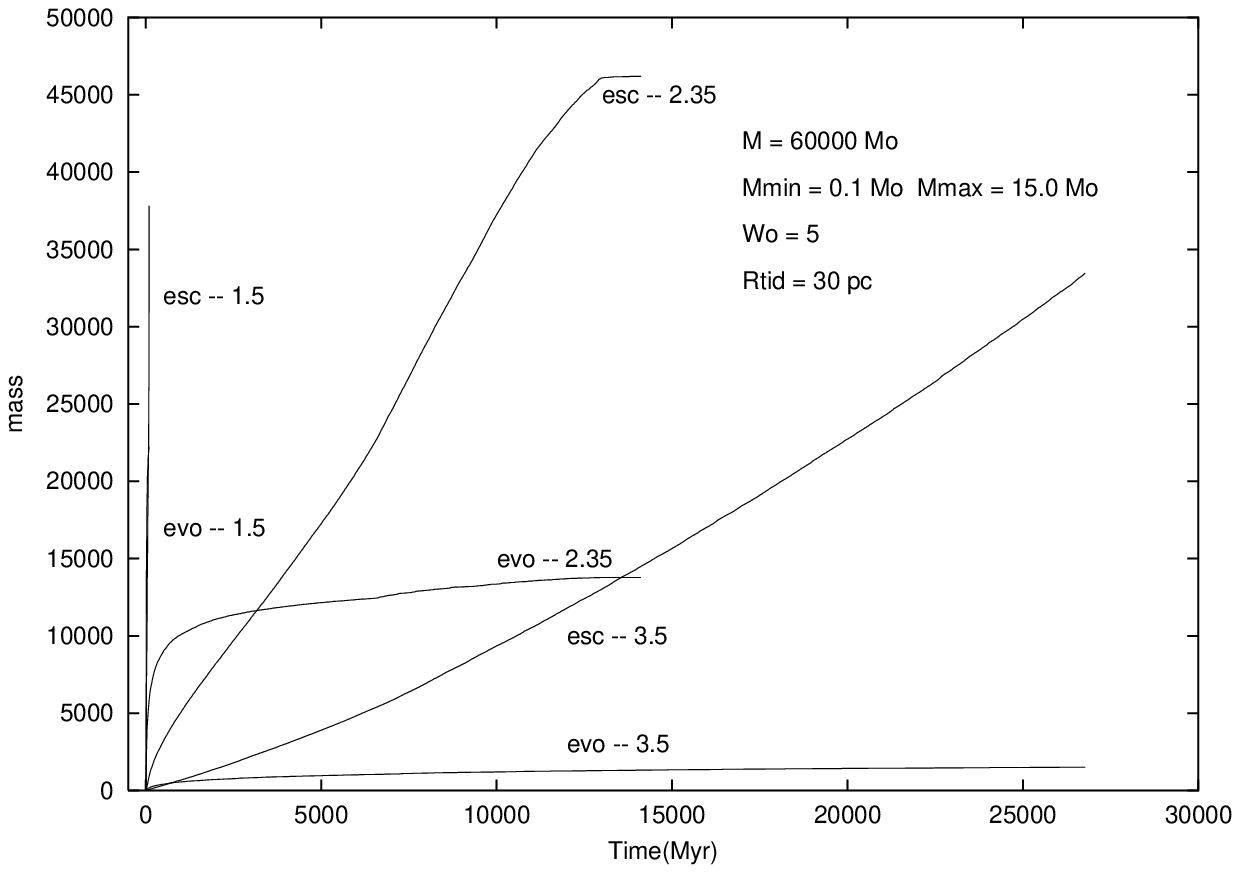}
\end{center}
\caption{Mass loss due to stellar evolution (evo) and tidal
stripping (esc) for models W515, W5235 and W535.}
\end{figure}
\begin{figure}
\epsfverbosetrue
\begin{center}
\leavevmode \epsfxsize=80mm \epsfysize=60mm
\epsfbox{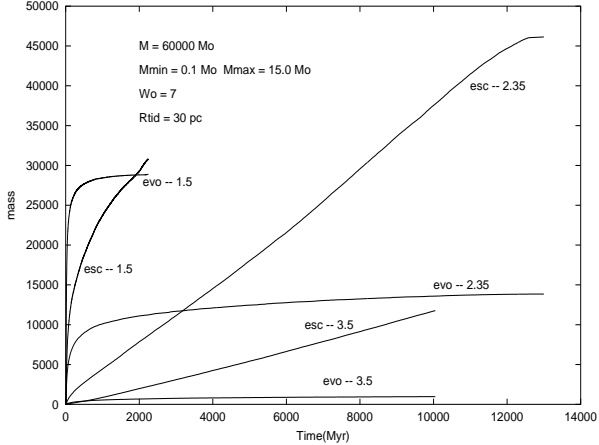}
\end{center}
\caption{Mass loss due to stellar evolution (evo) and tidal
stripping (esc) for models W715, W7235 and W735.}
\end{figure}
The final ratio of the amount of mass loss due to tidal stripping
to the amount of mass loss due to stellar evolution is smallest
for the shallow mass function ($\alpha = -1.5$). The steeper the mass
function, the higher the ratio. Mass loss connected with stellar
evolution dominates the initial phase of cluster evolution, as
could be expected. Then the rate of stellar evolution
substantially slows down and escape due to tidal stripping
takes over. During this phase of evolution the rate of mass loss
is nearly constant, and higher for shallower mass functions. Energy
carried away by stellar evolution events dominates the energy loss
due to tidal stripping, even though the tidal mass loss is higher
(see Figure 7). The higher the initial cluster concentration, the
smaller the absolute value of energy loss due to tidal stripping. For
a more concentrated cluster the overall mass loss due to tidal
stripping is smaller, and so the energy loss is smaller. It is
interesting to note, that for model W735 the energy loss due to
tidal stripping is positive (see Figure 7) during the second phase
of evolution. This means that stars with positive energy are
preferentially lost from the cluster, as in isolated systems. A
tidally limited system behaves like an isolated one, where strong
binary--single and binary--binary interactions are responsible for
escapers with a large positive energy. These relatively small numbers
of escapers energetically dominate the numerous escapers (due to tidal
stripping) with small negative energy.
\begin{figure}
\epsfverbosetrue
 \begin{center}
\leavevmode \epsfxsize=80mm \epsfysize=60mm \epsfbox{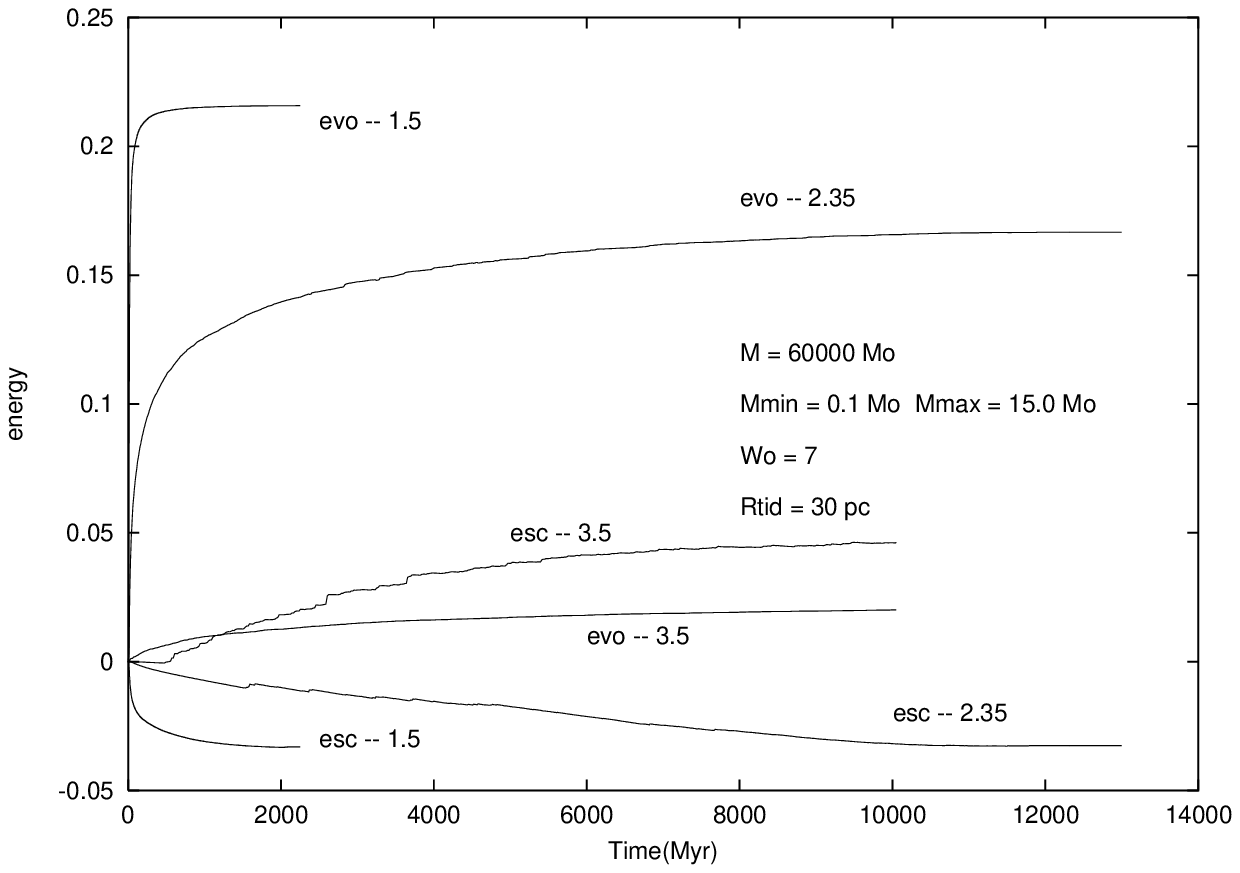}
\end{center}
\caption{Energy loss due to stellar evolution (evo) and tidal
striping (esc) for models W715, W7235 and W735.}
\end{figure}

The typical evolution of the central potential and Lagrangian radii
for the standard models is shown in Figures 8 and 9,
respectively. For clusters which are not disrupted due to strong
mass loss connected with stellar evolution of the most massive
stars, the three different phases of evolution can be clearly
distinguished. The first phase of violent mass loss due to stellar
evolution leads to overall cluster expansion and an increase of the
central potential. This phase is less pronounced in Figure 9,
because it is superposed on the shrinking of the tidal radius
(Lagrangian radii being calculated for smaller total mass). Because
of mass segregation the $1\%$ Lagrangian radius is decreasing.
The second phase is characterized by a decrease of the central
potential (except for models with low concentration and shallow
mass function, for which strong tidal stripping continuously
increases the central potential -- W325). The cluster undergoes core
collapse and behaves as an ordinary isolated system. The higher the
cluster concentration, the shorter this phase is. Then, in the third
phase, post--collapse evolution is superposed on effects of tidal
stripping. The central potential (on average) is continuously
increasing. The cluster contracts nearly homogeneously. The central
parts of the system show signs of gravothermal oscillations.
Family 1 models show the same features as discussed above.

\begin{figure}
\epsfverbosetrue
 \begin{center}
\leavevmode \epsfxsize=80mm \epsfysize=60mm
\epsfbox{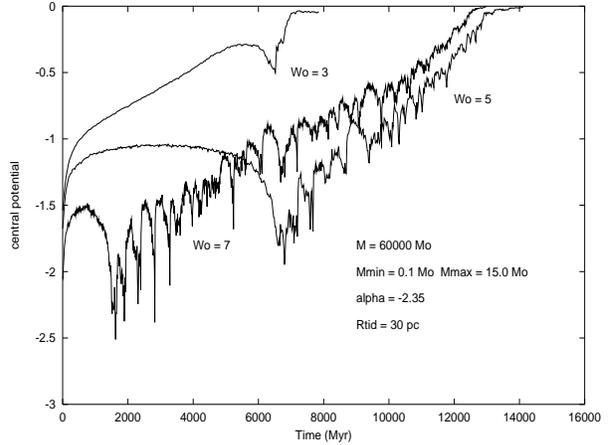}
\end{center}
\caption{Evolution of the central potential for models W715, W7235
and W735.}
\end{figure}
\begin{figure}
\epsfverbosetrue
 \begin{center}
\leavevmode \epsfxsize=80mm \epsfysize=60mm
\epsfbox{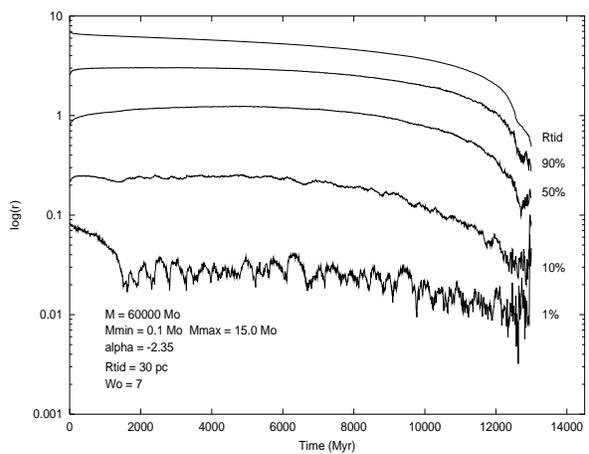}
\end{center}
\caption{Evolution of the Lagrangian radii for model W7235.}
\end{figure}
Generally, the agreement throughout the evolution between the results
presented here and these by AH for $N-$body models, TPZ for 2-D
Fokker--Planck models and by JNR for Monte Carlo models is rather
good. In all cases, the qualitative behaviour is identical, even
though the models of AH, TPZ and JNR have longer lifetimes than the
standard models.

\subsection{Anisotropy evolution.}

The degree of velocity anisotropy is measured by a quantity

\begin{equation}
\beta \equiv 2 - {2\sigma ^2_t \over {\sigma ^2_r}},
\end{equation}
where $\sigma_r$ and $\sigma_t$ are the radial and tangential
one--dimensional velocity dispersions, respectively. For isotropic
systems $\beta$ is equal to zero. Systems preferentially populated
by radial orbits have $\beta$ positive and systems preferentially
populated by tangential orbits have $\beta$ negative. The evolution
of anisotropy is presented in Figures 10 to 12 for models W515,
W725 and W735-4, respectively. As can be seen in Figure 10, for
systems which can not survive the violent initial mass loss,
the anisotropy stays close to zero. The system does not live long enough
to develop strong positive or negative anisotropy. These clusters which
undergo core collapse, are strongly concentrated and have a steep
mass function (W535, W7235), develop a small positive anisotropy in
the outer and middle parts of the system (see Figure 11). The
amount of anisotropy in the outer parts of the system is reduced
substantially by tidal stripping, as stars on radial orbits escape
preferentially. As tidal stripping
exposes deeper and deeper parts of the system, the anisotropy (for
large Lagrangian radii) gradually decreases and eventually becomes
slightly negative. Most stars on radial orbits were removed from
the systems and most stars on tangential orbits remained. At the
same time the anisotropy in the middle and inner parts of the system
stays close to zero. Finally, just before cluster disruption, the
anisotropy in the whole system becomes slightly positive again.
For model W735 the evolution of anisotropy proceeds in a very similar
way (at least long before the cluster disruption) as for isolated
clusters. Positive anisotropy develops throughout most of the
system. In the middle parts of the system it becomes nearly
constant just after core bounce (post--collapse evolution).
Only in the outer parts of the system does the anisotropy increase, showing
the development of the cluster halo. When tidal stripping becomes
important anisotropy will be reduced and it will behave as in the
model W725 discussed above. For models less strongly concentrated
and with a flatter mass function (W3235, W5235) the anisotropy in the
outer parts of the system very quickly becomes negative. Strong
stellar evolution and induced tidal stripping preferentially force
these stars to stay in the system, which are on tangential orbits. Just
before cluster disruption the anisotropy in the whole system again becomes
slightly positive.

For all models of Family 1 (see, for example, Figure 12) the evolution
of anisotropy for the outer parts of the system proceeds in a
similar way as for standard models W3235 and W5235. Family 1 models
contain more massive stars than the standard models. Therefore the
stronger mass loss very quickly forces the anisotropy in the outer parts
of the system to become negative. It stays negative until
cluster disruption, when it becomes slightly positive.
The anisotropy in the central parts of the system stays close to zero.
\begin{figure}
\epsfverbosetrue
 \begin{center}
\leavevmode \epsfxsize=80mm \epsfysize=60mm
\epsfbox{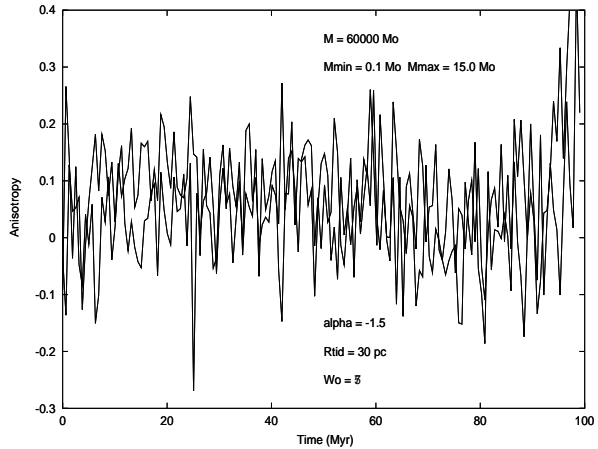}
\end{center}
\caption{Evolution of the anisotropy for 40\% and 99\% Lagrangian
radii for model W515.}
\end{figure}
\begin{figure}
\epsfverbosetrue
 \begin{center}
\leavevmode \epsfxsize=80mm \epsfysize=60mm
\epsfbox{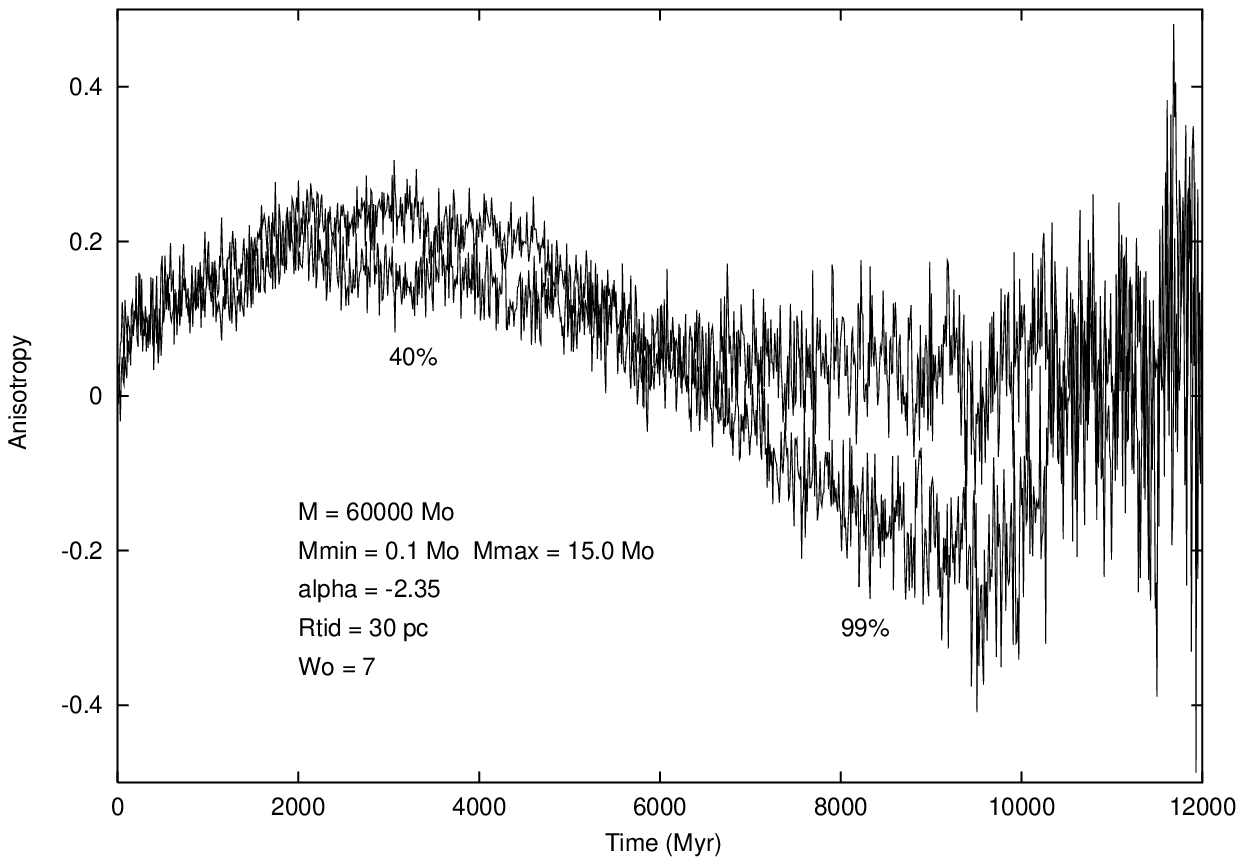}
\end{center}
\caption{Evolution of the anisotropy for 40\% and 99\% Lagrangian
radii for model W7235.}
\end{figure}
\begin{figure}
\epsfverbosetrue
 \begin{center}
\leavevmode \epsfxsize=80mm \epsfysize=60mm
\epsfbox{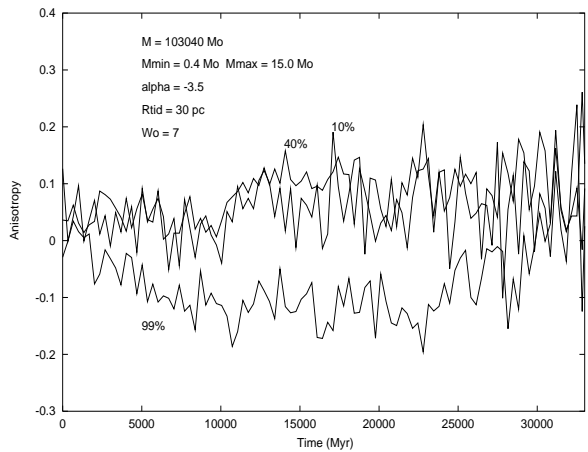}
\end{center}
\caption{Evolution of the anisotropy for 10\%, 40\% and 99\%
Lagrangian radii for model W735-4 (Family 1).}
\end{figure}

The anisotropy of the main--sequence stars shows very similar behaviour to
that discussed above. Mass segregation forces white dwarfs
and neutron stars to occupy the inner parts of the system  preferentially.
Therefore the  anisotropy for them is much more modest than for the
main--sequence stars, and stays close to zero. The evolution of
the anisotropy agrees qualitatively with the results obtained by Takahashi
(1997).

\subsection{Mass segregation.}

Three main effects may be expected to govern the evolution of the
mass function in models of the kind studied in this paper.
First, there is a period of violent mass loss due to evolution of
the most massive stars. It takes place mainly during
the first few hundred million years and its amplitude strongly
depends on the slope of the mass function. Secondly, there is the
process of rapid mass segregation, caused by two--body distant
encounters (relaxation). It takes place mainly during
core collapse and, as can be expected, is relatively
unaffected by the presence of a tidal field. Thirdly, there is the
effect of the tidal field itself, which becomes important after
core collapse or even earlier (for models with a shallow mass
function and with low concentration). Because the tides
preferentially remove stars from the outer parts of the system,
which (because of mass segregation) are mainly populated by
low--mass stars, one can expect that the mean mass should increase
as evolution proceeds, making allowance for stellar evolution.

The basic results are illustrated in Figures 13 to 15 for the total
average mass inside the $10\%$ and $50\%$ Lagrangian radii and the tidal
radius, and in Figures 16 to 18 for the overall average mass of
main--sequence stars and white dwarfs. The violent and strong mass
loss due to stellar evolution is characterized by an initial decrease
of the average mass. This is best seen in Figure 13 for model
W715. The initial average mass in this model is much larger than
for other models. It contains a relatively small number of low
mass stars and a large number of massive stars. Therefore the
evolution of the most massive stars will remove a substantial amount
of mass  from the system and, consequently, greatly lower the average
mass in the whole system. This behaviour is visible (with smaller
amplitude) in other models (with steeper mass function) as well, but
with one exception. The average mass inside the $10\%$ Lagrangian radius is
increasing instead of decreasing (see Figures 14 and 15). For these
models, there is enough time for mass segregation to force the most
massive stars (in this situation less massive main--sequence stars,
neutron stars and massive white dwarfs) to sink into the center and
increase the average mass there.

Family 1 models show a much stronger initial decrease of the average
mass than the standard models. This is connected with the
fact that they contain a much larger number of massive stars than the
standard models. The initial mass segregation is only visible for
model W735-4 for the 10\% Lagrangian radius.

For the standard models with
$\alpha = -2.35$ and $-3.5$, mass segregation substantially slows down
in the inner parts of the system after the core collapse. This is in
good agreement with results obtained by Giersz \& Heggie (1997) for
small $N$-body simulations, but the reason for that behaviour is
still unclear. During the post--collapse phase, for systems with a
steeper mass function, mass segregation proceeds further, but at a
smaller rate. The effect of tides manifests itself by a gradual increase
of the average mass. This increase, as expected, is fastest in the
halo and slowest in the core. There is one exception to this rule
models which are disrupted before core collapse show a nearly
constant rate of increase of the average mass inside the whole
system (see Figures 13 and 14). This is probably connected with
the fact, that for these disrupting systems stars are removed from
the whole body of the cluster.

Family 1 models which enter the long post--collapse evolution
phase show a steady, slow decrease of the average mass in the whole
system. The total initial average mass for these models is much
larger than for the standard models, and therefore stellar evolution
is important for a longer time and is competitive with tidal stripping.
\begin{figure}
 \epsfverbosetrue
  \begin{center}
 \leavevmode
\epsfxsize=80mm \epsfysize=60mm
 \epsfbox{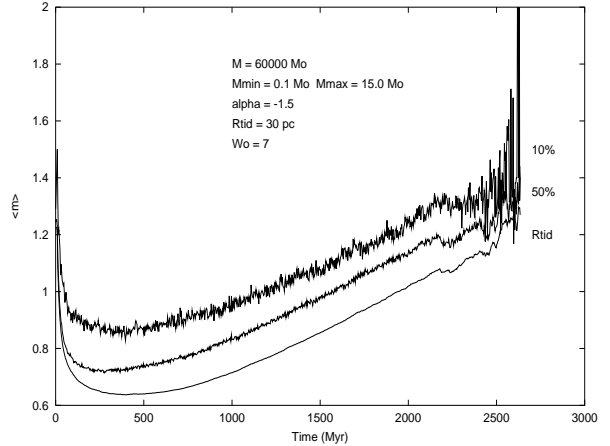}
\end{center}
\caption{Evolution of the average mass inside 10\%, 40\%
Lagrangian radii and $r_{tid}$ for model W715.}
\end{figure}
\begin{figure}
\epsfverbosetrue
 \begin{center}
\leavevmode \epsfxsize=80mm \epsfysize=60mm
\epsfbox{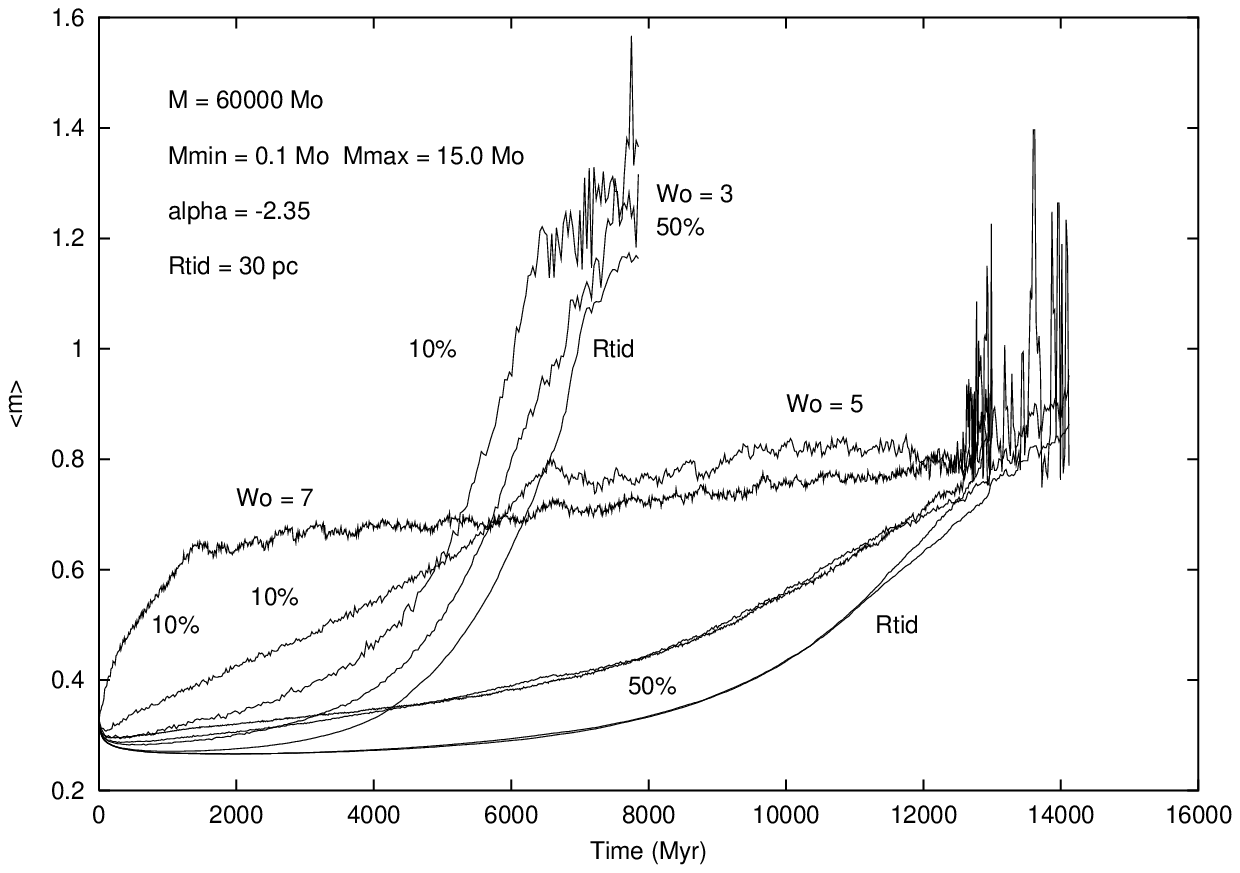}
\end{center}
\caption{Evolution of the average mass inside 10\%, 40\%
Lagrangian radii and $r_{tid}$ for models W3235, W5235 and W7235.}
\end{figure}
\begin{figure}
\epsfverbosetrue
 \begin{center}
\leavevmode \epsfxsize=80mm \epsfysize=60mm
\epsfbox{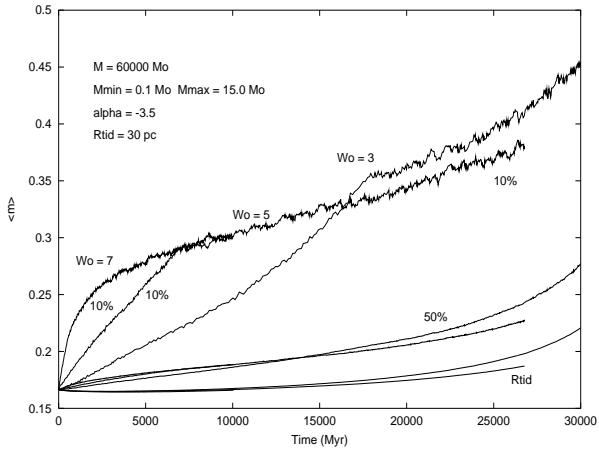}
\end{center}
\caption{Evolution of the average mass inside 10\%, 40\%
Lagrangian radii and $r_{tid}$ for models W335, W535 and W735.}
\end{figure}

The evolution of the overall average mass for main--sequence stars
and white dwarfs in the standard models shows basically the same
features as discussed above (see Figures 16 to 18). When less and
less massive main--sequence stars finish their evolution as less
and less massive white dwarfs, the average mass of white dwarfs
decreases with time. It seems, that in nearly the whole first Gyr
the rate of decrease of white dwarf average mass is practically
independent of the central concentration and the initial mass
function. The differences start to build up later and become most
evident close to the collapse time. This can be explained by the
fact that models with a steeper mass function contain a smaller
number of very massive main--sequence stars. Therefore the total
mass of white dwarfs is smaller for these models than for models
with a shallower mass function. When less and less massive
main--sequence stars finish their evolution, the newly created
less massive white dwarfs affect the average mass of white dwarfs
more strongly for the steeper mass function than for the shallower
one. At the time around core bounce the average mass of white
dwarfs increases, particularly for low concentration models with a
shallow mass function. Binaries start to be created mainly from
the most massive stars (neutron stars and white dwarfs) deep in
the core. In interacting with field stars these binaries remove
them from the system -- preferentially less massive white dwarfs.
At the time of core bounce less concentrated models with a
shallower mass function are on the verge of disruption. They
contain only a small number of stars, and therefore removal of
some low mass white dwarfs can lead to substantial changes of the
average mass.

For models which enter post--collapse evolution the average
mass of main--sequence stars behaves as expected. There is a small
initial decrease of the average mass, connected with stellar
evolution of the most massive stars. Then there is a period of the
gradual increase of the average mass due to tidal stripping and the
preferential removal of the least massive stars. At late
phases of cluster evolution the increase of average mass speeds
up. A different behaviour of the average mass of main--sequence
stars is observed for clusters which are disrupted just before
core collapse (W515 and W715). For these models removal of low
mass main--sequence stars is so strong (because of violent mass
loss due to stellar evolution of the most massive stars and
induced tidal stripping) that the average mass increases slightly
(see Figure 17 and 18). Then the average mass quickly decreases
because of stellar evolution and nearly homologous mass removal
from the system -- most stars are on radial orbits (see Figures 12
and 13).

Generally, Family 1 models show the same features as discussed
above. Only for models which enter the long post--collapse
evolution phase does the average main--sequence mass decrease slightly
instead of increasing. This is in agreement with the evolution of the
total average mass for Family 1 models discussed above.
\begin{figure}
\epsfverbosetrue
 \begin{center}
\leavevmode \epsfxsize=80mm \epsfysize=60mm
\epsfbox{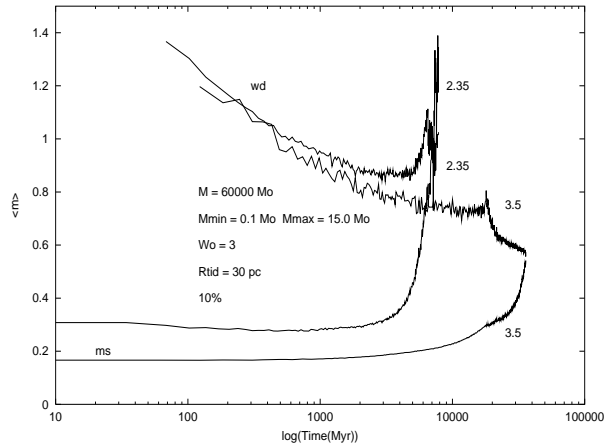}
\end{center}
\caption{Evolution of the average mass for 10\% Lagrangian radius
for models W3235, W335; $ms$ and $wd$ means mean--sequence stars
and white dwarfs, respectively.}
\end{figure}
\begin{figure}
\epsfverbosetrue
 \begin{center}
\leavevmode \epsfxsize=80mm \epsfysize=60mm
\epsfbox{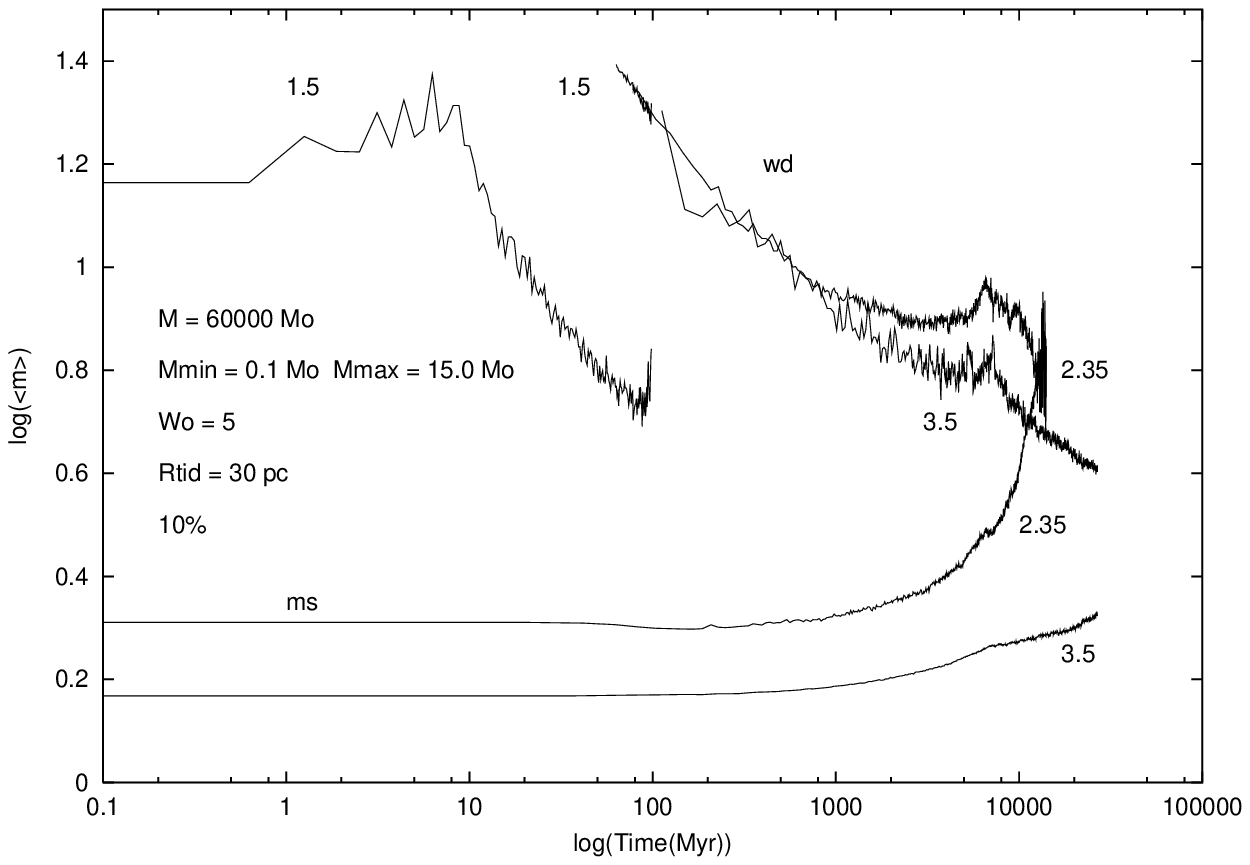}
\end{center}
\caption{Evolution of the average mass for 10\% Lagrangian radius
for models w515, W5235 and W535; $ms$ and $wd$ means
mean--sequence stars and white dwarfs, respectively.}
\end{figure}
\begin{figure}
\epsfverbosetrue
 \begin{center}
\leavevmode \epsfxsize=80mm \epsfysize=60mm
\epsfbox{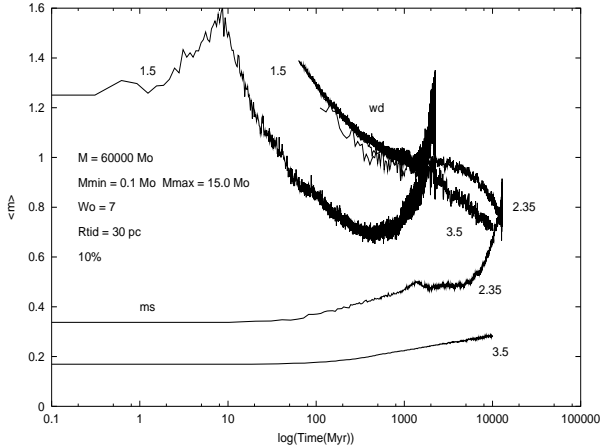}
\end{center}
\caption{Evolution of the average mass for 10\% Lagrangian radius
for models w715, W7235 and W735; $ms$ and $wd$ means
mean--sequence stars and white dwarfs, respectively.}
\end{figure}

\section{CONCLUSIONS.}
\medskip

This paper is continuation of Paper I, in which it was shown that
the Monte Carlo method is a robust scheme to study, in an
effective way, the evolution of a large $N$--body systems. The Monte
Carlo method describes in a proper way the graininess of the
gravitational field and the stochasticity of the real $N$--body
systems. It provides, in almost as much detail as $N$--body
simulations, information about the movement of any object in the
system. In that respect the Monte Carlo scheme can be regarded as a
method which lies between direct $N$--body and Fokker--Planck
models and combines most advantages of both these methods. The code
has been extended to include stellar evolution, multi--component
systems described by a power--law mass function, the tidal field of a
parent galaxy and generation of energy by binary--binary
interactions. This is the first major step in the direction of
simulating the evolution of a real globular cluster.

A small survey (similar to these presented by CW, AH, TPZ and JNR)
on the evolution of globular clusters in the Galactic tidal field
was carried out. It was shown that the results obtained are in
qualitative agreement with these presented by CW, AH, TPZ and JNR.
Particularly good agreement is obtained with AH's $N$-body
simulations. JNR's Monte Carlo results, mainly for strongly
concentrated models, disagree with the other models. The Monte
Carlo scheme discussed in this paper has some problems with low
concentration models and these with a flat mass function. It seems
that the collapse times for these models are too short, in
comparison with the results of the other methods. However, these
models are problematic for all methods; for example, the $N$--body
method has problems with proper time scaling. All standard models,
for which mass loss due to violent stellar evolution of the most
massive stars does not induce quick cluster disruption, evolve in
a very similar way. The rate of mass loss, the evolution of the
central potential, and the evolution of the average mass, do not
depend much on the central concentration of the system. They
depend strongly on the index of the mass function. Models of
Family 1, on the contrary, show dependence on the initial
concentration, as well. The very high initial mass loss across the
tidal boundary, connected with the evolution of the most massive
stars (for models of Family 1, there are more massive stars than
for standard models) forces to substantial changes in the
structure of the system and in consequence different evolution of
the total mass, anisotropy, etc. Models which are quickly
disrupted show only small signs of mass segregation. Models with
larger central concentration survive the phase of rapid mass loss
and then undergo core collapse and subsequent post--collapse
expansion in a manner similar to isolated models. The expansion
phase is eventually reversed when tidal limitation becomes
important. As in isolated models, mass segregation substantially
slows down by the end of core collapse. After core bounce there is
a substantial increase in the mean mass in the middle and outer
parts of the system, caused by the preferential escape of stars of
low mass and by tidal effects. Standard models, which are not
quickly disrupted, show a modest initial build up of anisotropy in
the outer parts of the system. As tidal stripping exposes the
inner parts of the system, the anisotropy gradually decreases and
eventually becomes slightly negative. The central part of the
system stays nearly isotropic. From the very beginning models of
Family 1 develop a modest negative anisotropy in the outer parts
of the system. It stays negative until the time of cluster
disruption, when it becomes slightly positive (during cluster
disruption most stars are on radial orbits).

In order to perform simulations of real globular clusters several
additional physical effects have to be included into the code. The
tidal shock heating of the cluster due to passages through the
Galactic disk, interaction with the bulge, shock--induced
relaxation, primordial binaries, physical collisions between
single stars and binaries are some of them. Inclusion of all these
processes does not pose a fundamental theoretical or technical
challenge. It will allow us to perform detailed comparison between
simulations and the observed properties of globular clusters, and will
also help to understand the conditions of globular cluster formation and
explain how peculiar objects observed in clusters can be formed.
These kinds of simulations will also help us to introduce, in a proper
way, into future $N$--body simulations all the necessary processes
required to simulate evolution of real globular clusters on a
star--by--star basis from their birth to their death.
\bigskip
\bigskip

{\parindent=0pt {\bf Acknowledgments} I would like to thank
Douglas C. Heggie and Rainer Spurzem for stimulating discussions,
comments and suggestions. I also thank DCH for making the
$N$--body results of standard model simulations available. This
work was partly supported by the Polish National Committee for
Scientific Research under grant 2--P03D--022--12.}

\bigskip
\bigskip
\bigskip

\bsp

\end{document}